\documentclass[aps,showpacs,amsmath,dvips,10.50pt,showkeys,nofootinbib]{revtex4}

\usepackage{graphicx}   
\usepackage{color}      
\usepackage{braket}
\usepackage{tikz}
\usepackage{pgfplots}
\usepackage{orcidlink}

\def \be  {\begin{equation}}
\def \ee  {\end{equation}}
\def \ee  {\end{equation}}
\def \bea {\begin{eqnarray}}
\def \eea {\end{eqnarray}}

\begin{document}

\title{Non-extensive Effects on the QCD Equation of State and Fluctuations of Conserved Charges within Polyakov Quark Meson Model}

\author{Abdel Magied DIAB \orcidlink{0000-0002-5603-1952}}
\email{a.diab@eng.mti.edu.eg}
\affiliation{Faculty of Engineering, Modern University for Technology and Information (MTI), 11571 Cairo, Egypt.}

\date{\today}
\begin{abstract} 
The influence of non-extensive Tsallis statistics on the hadron phase structure has been investigated using the Polyakov-quark-meson (PQM) model. The analysis examines the non-extensive effects on the temperature dependence of PQM order parameters, thermodynamic quantities related to the QCD equation of state, and fluctuations of conserved charges at varying chemical potentials. The results show that non-extensive effects have the most significant deviations near the crossover region. The pseudo-critical temperature $T_{\chi}(\mu_B)$ is not a universal constant and decreases with increasing non-extensive $q$ parameter. The chiral phase diagram of the PQM model indicates a decrease in the behavior of the ($T_{\chi}-\mu_B$) plane with increasing non-extensive $q$ parameter. The PQM model exhibits good qualitative agreement with lattice QCD calculations. Moreover, these findings suggest the existence of a Tsallis limit, which serves as an alternative to the Stefan-Boltzmann (SB) limit for the massless ideal gas. The critical endpoint (CEP) exhibits lower temperature but higher chemical potential with increasing non-extensive $q$ parameter. Overall, this study highlights the importance of non-extensive Tsallis statistics in characterizing the quark-hadron phase structure of the PQM model and contributes to a deeper understanding of non-extensive effects in the quark-hadron phase transition.
\end{abstract}

\pacs{11.30.Rd, 11.10.Wx, 12.39.Fe, 02.60.-x, 12.40.Ee.}
\keywords{Chiral symmetries, Chiral transition, Chiral Lagrangian, Numerical approximation and analysis,  Statistical models.}
\maketitle

\section{Introduction \label{intro}}
Power-law distributions have been extensively explored across different scientific disciplines, including the well-known Tsallis distribution  \cite{tsallis1988possible}. The Tsallis distribution, which is an extension of the Boltzmann-Gibbs (BG) distribution with an additional parameter \cite{Tsallis:2009zex}, has gained significant attention in the past few decades. It has been widely employed to analyze various phenomena  \cite{Wilk:1999dr,Rybczynski:2014cha,Lavagno:2010xu,Lavagno:2012yn,Trevisan:2013qoa,Gervino:2013uoa,Lavagno:2013qca,Frigori:2014fla,Rozynek:2015iea,Ishihara:2014txa,Ishihara:2017txj,Ishihara:2018yxp,Ishihara:2019ran,Rozynek:2019qzq,Rahaman:2021xqv,Zhao:2020xob,Islam:2023zpl,Rozynek:2015zca,Rozynek:2009zh,Wilk:2006dv}, among many others such as astrophysics \cite{Tsallis:2012js, Lavagno:2011zm} and optical lattice \cite{Douglas:2006zz}. For instance, the Tsallis distribution has been used to describe the momentum distribution observed in high-energy collision experiments. Consequently, this framework has been widely accepted to analyze transverse momentum distributions observed in high-energy collisions. Notably, the PHENIX and STAR collaborations at the Relativistic Heavy Ion Collider (RHIC) in BNL \cite{STAR:2006nmo, PHENIX:2011rvu}, as well as the ALICE, ATLAS, and CMS collaborations at the Large Hadron Collider (LHC) in CERN \cite{ALICE:2011gmo,ALICE:2014juv,CMS:2010tjh,ATLAS:2010jvh}, have been at the forefront of utilizing the Tsallis distribution in experimental particle physics. As a result of the small size and rapid evolution of the Quark-Gluon Plasma (QGP), it exhibits a non-uniform distribution and lacks the global equilibrium. Consequently, certain quantities within the system display non-extensive behavior, characterized by power-law distributions instead of exponential distributions.

It is worth noting that BG statistics \cite{huang2009introduction, dalfovo1999theory} are commonly employed to describe the  quantum chromodynamics (QCD) phase transition. However, it is essential to recognize that BG statistics can only be applied to systems in equilibrium and within the thermodynamic limit. In relativistic heavy-ion collisions, the QGP produced undergoes strong intrinsic fluctuations and exhibits long-range correlations \cite{Sharma_2019,Bhattacharyya_2018}. The size of the QGP is sufficiently small, and its evolution is rapid. Consequently, this system is far from being equilibrium. It is not appropriate to apply the BG statistics to evaluate the energies and numbers of particles in such a system exhibits long-range correlations \cite{Stephanov:1998dy}. It is evident that the introduction of the non-extensive $q$ parameter in the PQM model plays a role in capturing the deviations from standard Boltzmann-Gibbs (BG) statistics  in equilibrium to non-equilibrium states. The studying of the phase structure of QCD matter, non-extensive statistics were first proposed by C. Tsallis in \cite{tsallis1988possible}, where non-extensivity is understood in the thermodynamic sense. The so-called Tsallis statistics propose the entropy as $S_q = \mbox{Tr}(\hat{\rho} - \hat{\rho}^q)/(q-1)$ and the density matrix as $\hat{\rho} = \exp_q \big(-E/T\big)/Z_q$, where $Z_q$ is the corresponding partition function. Here, $q$ is a real number that characterizes the non-extensive factors of such a system. The Tsallis distribution can be characterized by two essential parameters: the non-extensive $q$ parameter, which controls the transition between power-law and exponential behavior, and the Tsallis temperature $T$. Additional parameters such as mass and chemical potentials can be also included to further describe the distribution. When the $q$-parameter exceeds one, the Tsallis distribution exhibits a power-law behavior, whereas it converges to the conventional BG distribution as the q-parameter approaches one.

To address the limitations of lattice QCD calculations in dealing with non-extensive Tsallis statistics, researchers utilize various effective models to explore the phase diagram of QCD theoretically. These models include chiral perturbation theory \cite{Espriu:2020dge}, Dyson-Schwinger Equations (DSEs) \cite{Zhao:2019ruc,Fischer:2018sdj,Ayala:2011vs},  based on density operator \cite{Ishihara:2018yxp}, hadron resonance gas and statistical models \cite{Kyan:2022eqp,Parvan:2023jnx}, Nambu-Jona-Lasinio model and the Polyakov Nambu-Jona-Lasinio (PNJL) model \cite{Rozynek:2015zca,Zhao:2020wks,Zhao:2020xob,Zhao:2022ccj,Zhao:2023xpj}, linear sigma model and Polyakov linear sigma model \cite{Shen:2017etj,Ishihara:2016moh,Ishihara:2018cgj}.

The PQM model goes beyond merely predicting the order of the chiral phase transition by combining two fundamental non-perturbative aspects of low-energy QCD: confinement and spontaneous chiral symmetry breaking in a simple approach. It provides a good interpretation of lattice QCD simulations at zero baryon density and finite temperature. Most effective models, including the PQM model, describe the hadron-quark phase transition based on quark degrees of freedom. The degrees of freedom in  $U(N_f)_r \times U(N_f)_\ell$ PQM model encompass chiral meson states, where the number of quark flavors is $N_f$. In the vacuum, the latter serve as (pseudo-) Goldstone bosons resulting from chiral symmetry breaking \cite{Roder:2003uz}. Consequently, thePQM model can also be viewed as an effective theory for low-energy degrees of freedom of the QCD. At high temperatures, chiral symmetry undergoes restoration, causing the chiral partners among meson states to become degenerate in mass. It is particularly suited to describe the chiral phase structure of meson states \cite{Tawfik:2014gga,Tawfik:2019rdd}. In PQM model, we can not involve the degrees of freedom of hadrons directly, but this effective low-energy QCD model captures the confinement phase by suppressing the excitations of the quark states \cite{Fu:2007xc}. In this context, the term "confinement" refers to the statistical suppression of the propagation of color-triplet quarks. Similarly, the chiral symmetry is explicitly breaking due to the nonzero current masses of quarks. However, the chiral condensate remains a valuable indicator of the chiral crossover. The validity of the PQM model has been confirmed through a series of works that involved comparisons with Lattice QCD simulations, for instance, \cite{Tawfik:2014uka,Tawfik:2016edq,AbdelAalDiab:2018hrx, Tawfik:2019kaz}, including the influence of the isospin asymmetry \cite{Tawfik:2019tkp} and chiral magnetic effect \cite{Ezzelarab:2015tya,Tawfik:2016lih,Tawfik:2016gye,Tawfik:2016ihn,Tawfik:2017cdx,Tawfik:2021eeb}. Moreover, the Polyakov loop extended quark meson model containing  vector and axial vector mesons \cite{Kovacs:2016juc} and large $N_c$-limit \cite{Kovacs:2022zcl}.

The motivation behind this paper is to examine the effects of non-extensive Tsallis statistics on the chiral phase structure, thermodynamic properties, and fluctuations of conserved charges in QCD matter at finite temperature $T$ and chemical potentials $\mu$. To accomplish this, we extend the PQM model by incorporating non-extensive Tsallis statistics, allowing us to compare the results with those obtained using BG statistics.

The paper is organized as follows: Sec. \ref{PQMModel} presents the formalism of the PQM model incorporating non-extensive Tsallis statistics. In Sec. \ref{Results}, we analyze the influence of the non-extensive $q$ parameter on PQM Order Parameters (Subsec. \ref{OrderParameter}), explore the chiral phase diagram for different selections of the non-extensive $q$ parameter, especially the location of CEP due to the non-extensive effects  (Subsec. \ref{subtCond}), investigate the impact on various thermodynamic quantities with different non-extensive $q$ parameters (Subsec. \ref{ThermoQ}), and examine the effects on the fluctuations and correlations of conserved charges (Subsec. \ref{Fluctuations}). Finally, Sec. \ref{conclusion} provides a brief summary and conclusion. 

\section{Non-extensive Tsallis Statistics in the Polyakov Quark Meson Model \label{PQMModel}}
The Lagrangian of the PQM model can be incorporated from chiral Lagrangian of $N_f=3$ quark flavors and the Polyakov-loop potential as
\bea
\mathcal{L}_{PQM} &=&  \mathcal{L}_{\mbox{chiral}} -\mathbf{\mathcal{U}}(\phi, \bar{\phi}, T) = \mathcal{L}_{\bar{\psi}\psi}+\mathcal{L}_{\mathcal{M}}-\mathbf{\mathcal{U}}(\phi, \bar{\phi}, T). \label{PQM_Eq}
\eea
The first term of rhs of Eq. (\ref{PQM_Eq}), gives the contributions of quarks (fermions) with three-color degrees-of-freedom $N_c=3$, 
\bea
\mathcal{L}_{\bar{\psi}\psi} &=& \sum_f \overline{\psi}_f(i\gamma^{\mu} D_{\mu}-g\,T_a(\sigma_a + i \gamma_5 \pi_a))\psi_f, \label{lfermion} 
\eea
where $D_{\mu},\; \mu,\; g, \gamma^{\mu}$ and $\psi$ are defined as the covariant derivative, Lorentz index, Yukawa coupling constant, chiral spinors and Dirac spinor fields, respectively. The index $f$ refers to quark flavors as $f=[(l=u, d), s]$. The second term in rhs of Eq. (\ref{PQM_Eq}) shows the contributions of mesons (bosons), 
\bea
\mathcal{L}_{\mathcal{M}} &=&\mathrm{Tr}\Big[\partial_{\nu} \mathcal{M}^{\dag}\partial^{\nu} \mathcal{M}-m^2
\mathcal{M}^{\dag} \mathcal{M}\Big]-\lambda_1\,\Big(\mathrm{Tr}\,\Big[\mathcal{M}^{\dag} \mathcal{M}\Big]\Big)^2 - \lambda_2\,\mathrm{Tr}\Big[\mathcal{M}^{\dag} \mathcal{M}\Big]^2+  \nonumber \\
&+& c\Big(\mathrm{Det}[\mathcal{M}]+\mathrm{Det}[\mathcal{M}^{\dag}]\Big) +\mathrm{Tr}\Big[H(\mathcal{M}+\mathcal{M}^{\dag})\Big],  \hspace*{6mm} \label{lmeson} 
\end{eqnarray}
where $\mathcal{M}$ is a complex $(3\times3)$ matrix which can be given for nonet meson states as, $\bar{\mathcal{M}} =  \sum_{a=0}^{N_f^2 - 1} T_a ( \bar{\sigma_a}+i  \bar{\pi_a})$. In U($3$) algebra, the generator operator $T_a= \hat{\lambda}_a/2$ can be estimated from Gell-Mann matrices $\hat{\lambda}_a$ \cite{Weinberg:1972kfs} with the indices $a=0,\cdots,\,8$. The parameters, $m^2$, $h_l$, $h_s$, $\lambda_1$, $\lambda_2$, and $c$ are given in dependence on the mass of the sigma meson $m_\sigma$ \cite{Schaefer:2008hk}. The present work used the values of these parameters at $m_\sigma=800~$MeV \cite{Schaefer:2008hk}. The chiral symmetry is explicitly broken by ($3\times 3$) matrix $H= T_a h_a$, the explicitly symmetry breaking terms $h_a$ can be determined from the partially conserved axial current (PCAC) relations \cite{Lenaghan:2000ey,Tawfik:2019tkp}.

The third term in rhs of Eq. (\ref{PQM_Eq}), $\mathcal{U}(\phi, \bar{\phi},T)$ represents the potential of the Polyakov-loop variables, this potential incorporates the color gluonic degrees-of-freedom and dynamics of the quark-gluon interactions into the chiral Lagrangian. The potential is obtained by calculating the thermal expectation value of a color-traced Wilson-loop in the temporal space as $ \phi =  (\mathrm{Tr}_c\, \mathcal{P})/N_c$, and its hermitian conjugate  $\bar{\phi} = (\mathrm{Tr}_c\,  \mathcal{P}^{\dag})/N_c $, where $\mathcal{P}$ is the matrix the Polyakov loop operator in SU$(N_c)$ gauge group as,
\bea
\mathcal{P}(\vec{x}) = P \exp\Big({i\int_0^{1/T} d\tau A_0 (\vec{x}, \tau)}\Big), 
\eea
where $A_0, P$ are the temporal vector field and path ordering, respectively. The expectation value of the Polyakov loops is finite and approach to the unity (deconfined phase)  at high temperatures and it vanishes or small values at low tempertures (confined phase) \cite{Schaefer:2009ui}. Therefore, this order parameter is a good indicator of the (de)confinement  transition phase transition. Various expressions have been suggested for the potentials of the Polyakov loop variables, which are generally considered to represent the QCD symmetries in pure-gauge theory. For further information, \cite{Ratti:2005jh,Roessner:2006xn,Schaefer:2007pw,Fukushima:2008wg, Lo:2013hla}. In the present work, we utilize the effective polynomial expression of Polyakov loop variables, Eq.(\ref{Ppot1}), proposed in Ref.\cite{Ratti:2005jh,Pisarski:2000eq,Scavenius:2002ru} and compared with Fukushima potential, Eq.(\ref{FUKU}), as in Ref. \cite{Fukushima:2008wg}. Both of these expressions incorporate higher-order Polyakov-loop variables and are derived from strong coupling simulations:
\begin{itemize} 
\item The polynomial form of the  Polyakov-loop potential (Pot. I) \cite{Ratti:2005jh,Pisarski:2000eq,Scavenius:2002ru} is given by,
\bea
\mathbf{\mathcal{U}}_{\mathrm{Poly}}(\phi, \bar{\phi}, T) &= & T^4 \Big(-\frac{b_2(T)}{2}\, \bar{\phi}\phi - \frac{b_3}{6}\,  (\bar{\phi}^3+\phi^3) + \frac{b_4}{4}\,  (\bar{\phi}\phi)^2\Big), \label{Ppot1}
\eea
with parameters are a temperature dependent \cite{Ratti:2005jh,Pisarski:2000eq,Scavenius:2002ru} as
$b_2(T) = a_0 +a_1 (T_0/T)+a_2 (T_0/T)^2+ a_3 (T_0/T)^3 $.
\item While the Fukushima version from Polyakov-loop potential  (Pot. II) \cite{Fukushima:2008wg} reads as,
\bea
 \mathbf{\mathcal{U}}_{\mathrm{Fuku}}(\phi, \bar{\phi}, T) & = & - b \;T \left[54\, \phi \,\bar{\phi} \;\exp(-a/T) + \ln(1-6 \phi \bar{\phi} - 3(\phi \bar{\phi})^2+4 (\phi^3 + \bar{\phi}^3))\right], \label{FUKU}
\eea 
\end{itemize}
where $a$ and $b$ are fixed parameters \cite{Fukushima:2008wg}. The partition function can be formulated for a spatially homogeneous system in thermal equilibrium at a finite temperature $T$ and quark chemical potentials $\mu_f$. The exchange in particle and antiparticle  can be governed by the grand canonical partition function. By performing a path integral over the (anti)quark, and meson fields, the thermodynamic grand-canonical potential can be deduced as, 
\bea
\Omega(T, \mu_f) &=& \frac{-T\;\cdot \ln{\left[\mathcal{Z}\right]}}{V} 
= \frac{-T}{V} \ln \mathrm{Tr}\Big[\exp{\Big(\frac{-1}{T} \int d^3x (\mathcal{H})-\mu_f (\psi^\dagger\psi \Big)}\Big],
\label{potential}
\eea  
where $\mu_f$ refers to the chemical potentials of the quark flavors $f=[l=(u, d), s]$, which are associated to the  conserved charge numbers of - for instance - baryon ($B$), electric charge ($Q$) and strangeness ($S$) of each quark flavor, 
\bea
\mu_u = \frac{\mu_B}{3} + \frac{2 \mu_Q}{3}, \qquad   \qquad
\mu_d = \frac{\mu_B}{3} - \frac{\mu_Q}{3},  \qquad   \qquad
\mu_s = \frac{\mu_B}{3} - \frac{\mu_Q}{3} -\mu_S. \label{muf-eqs}
\eea
Using the mean field approximation (MFA), the PQM thermodynamic potential $\Omega$ can be obtained as,
\bea
\Omega(T, \mu_f) &=& U(\sigma_l,\, \sigma_s)+\Omega_{\bar{\psi}\psi} (T,\,\mu_f) +\mathbf{\mathcal{U}}(\phi, \bar{\phi}, T). \label{PQMPot}
\eea
More details about the structure of the PQM model can be found in Refs. \cite{Tawfik:2014gga,Tawfik:2021eeb}. The first term in rhs of Eq. (\ref{PQMPot}) gives the potential of the contribution of mesons based on non-strange $\sigma_l$  and strange $\sigma_s$ condensate as,
\bea
U(\sigma_l,\, \sigma_s) &=&- h_l \, \sigma_l - h_s\, \sigma_s +\frac{m^2}{2} \Big(\sigma_l^2+\sigma_s^2\Big) -\frac{c}{2\sqrt{2}} \sigma_l^2\, \sigma_s +\frac{\lambda_1}{4} \Big(\sigma_l^2+\sigma_s^2\Big)^2+\frac{\lambda_2}{4}\Big(\frac{\sigma_l}{4}+\sigma_s^4\Big), \label{sigmaPot}
\eea
where the parameters and coupling constants are listed in Eq. (\ref{sigmaPot}) are fixed  for $m_\sigma=800~$MeV \cite{Schaefer:2008hk}. The second term in rhs of Eq. (\ref{PQMPot}) shows the potential of quark and anti-quark contributions in finite temperature $T$ and quark chemical potential $\mu_f$ as \cite{Fukushima:2003fw},
\bea
\Omega_{\bar{\psi}\psi}(T, \mu _f)&=& -2N_c\,N_f \int \frac{d^3k}{(2\pi)^3}\Big\{\; E_f + \frac{T}{N_c}\;\Big(\ln{[\mathcal{F_D^{(+)}}(T, \; \mu_f)]} + \ln{[\mathcal{F_D^{(-)}}(T, \; \mu_f) ]} \Big)\Big\}, \label{qaurkPot}
\eea  
where $\mathcal{F_D^{(+)}}(T, \; \mu_f)$ and $\mathcal{F_D^{(-)}}(T, \; \mu_f)$ are defined as the Fermi-Dirac distribution functions in the present of the dynamics of the quark-gluon interactions by Polyakov loop variables - in a widely used notations - as,
\bea
\mathcal{F_D^{(+)}}(T, \; \mu_f) &=&   1+ 3\left(\phi\,+\bar{\phi}\;e^{-\frac{E_f^{(+)}}{T}}\right) e^{-\frac{E_f^{(+)}}{T}}+e^{-3 \frac{E_f^{(+)}}{T}}  , \label{Modfermi}
\eea
where $E_f^{(\pm)}=E_f\mp \mu_f$ are the energy-momentum dispersion relations, in which the upper sign is applied for quarks and the lower sign for antiquarks. The quark-flavor energy-momentum dispersion relations are given as  $E_f=(k^2+m_f^2)^{1/2}$, where $m_f$ is the constituent quark masses for non-strange quark sectors $m_l= g\sigma_l/2$ and strange quark sector $m_s = g\sigma_s/\sqrt{2}$.  One can estimate that the term $\mathcal{F_D^{(-)}}(T, \; \mu_f)$, by replacing $E_f^{(+)} $ with $E_f^{(-)}$ and the the Polyakov-loop variable $\phi$ with its conjugate $\bar{\phi}$ and vice versa.  After estimating the PQM thermodynamic potential in MFA, one can determine the thermodynamic quantities that describe the chiral phase structure of quark-hadron matter at finite temperatures and finite chemical potentials. These quantities can be obtained by examining the behaviors of the PQM order parameters.

In mean field approximation, one can estimate the quark condensates $\bar{\sigma}_l,\;\bar{\sigma}_s$ and the expectation values of the Polyakov-loop variables $\bar{\phi},\;\bar{\bar{\phi}}$ by searching the global minimization of the real part of the he PQM thermodynamic potential $\mathcal{R} \;[\Omega (T, \, \mu)]$, Eq. (\ref{PQMPot}), with respect to the associated order parameter, at a saddle point,

\bea
\frac{\partial\Omega}{\partial\sigma_l}= \frac{\partial\Omega}{\partial\sigma_s} = \frac{\partial\Omega}{\partial\phi} = \frac{\partial\Omega}{\partial\bar{\phi}}\Bigg\vert_{\sigma_l=\bar{\sigma}_l,\sigma_s=\bar{\sigma}_s,\phi=\bar{\phi},\bar{\phi}=\bar{\bar{\phi}}} = 0. \label{saddleEq}
\eea 

However, the Polyakov-loop is a complex matrix. Thus, the PQM thermodynamic potential, Eq. (\ref{PQMPot}), is a functional of complex variables. The physical quantities of the imaginary part of this potential vanish \cite{Sasaki:2006ww}. The PQM order parameters are given by minimizing the real part of the thermodynamic potential globally at a saddle point. They are determined by solving the four equations as a complete set of equations. The effective action turns complex, and only the saddle point can be defined, which is generally not a global minimum. However, depending on the directions of approach to the saddle point. Where the gap equations are well-defined, the behaviors of PQM order parameters are analyzed.

The non-extensive Tsallis statistics is characterized by the substitution of the conventional exponential factors with their $q$-exponential counterparts as, 
\bea
f(x) &=& \exp{(x)}  \longrightarrow  f_q(x) = \exp_q{(x)}= \Big(1+(1-q)x\Big)^{\frac{1}{1-q}}, \\
g(x) &=&  \ln{(x)}  \longrightarrow g_q(x) =  \ln_q{(x)} = \frac{x^{1-q} -1}{1-q}.
\eea
The non-extensive $q$ parameter indicates all possible factors does not contain in BG Statistics. As $q$ approaches unity, the non-extensive Tsallis statistics converges to the Boltzmann-Gibbs statistics. For this case study, we will make two assumptions. Firstly, we will not directly incorporate non-extensive effects on the pure Yang-Mills sector. Consequently, the Polyakov-loop potential remains unchanged, but the non-extensive impacts on Polyakov-loop variables are indirectly included through the saddle point, Eq. (\ref{saddleEq}). Secondly, the fixed parameters selected for the present approach remain unchanged. At zero temperature and zero chemical potential, these parameters are fitted to match the observed calculations. We can rely on the assumption that these parameters can be utilized at finite temperature and quark chemical potential. Thus,the potential of quark and anti-quark contributions, Eq. (\ref{qaurkPot}), can be written within non-extensive $q$ parameter as,
\bea
\Omega_{q}(T, \mu _f)&=& -2N_c\,N_f \int \frac{d^3k}{(2\pi)^3}\Big\{\; E_f + \frac{T}{N_c}\;\Big(\ln_q{[\mathcal{F_D^{(+)}}_q(T, \; \mu_f)]} + \ln_q{[\mathcal{F_D^{(-)}}_q(T, \; \mu_f) ]} \Big)\Big\}, \label{qaurkPotq}
\eea
where
\bea
\mathcal{F_{D}^{(+)}}_q(T, \; \mu_f) &=&   1+ 3\left(\phi\,+\bar{\phi}\;e_q^{-\frac{E_f^{(+)}}{T}}\right) e_q^{-\frac{E_f^{(+)}}{T}}+e_q^{-3 \frac{E_f^{(+)}}{T}}. \label{Modfermiq}
\eea

The effective pattern has a complex structure and scaling properties arising from self-energy interactions, characterized as a so-called thermofractal system \cite{Deppman:2016fxs,Deppman:2017fkq}. This system can be described by the Yang-Mills theory \cite{Callan:1970yg,Symanzik:1970rt}. The thermofractal structure in its thermodynamic distributions can be be follow from the non-extensive thermodynamics proposed by Tsallis \cite{tsallis1988possible}. The thermofractal system with quantum fields was proposed in Ref. \cite{Deppman:2019yno}. However, the thermofractal system leads to a recursive equation that provides the running-coupling constant as given in \cite{Deppman:2019yno}. 
\bea
g(\epsilon_0) = \prod_{i=1}^2 \Big( 1- (q-1) \frac{\epsilon_i}{\lambda}\Big)^{\frac{1}{q-1}},
\eea  
where  $\lambda$ and $\epsilon_i$ are defined as a scale parameter and energy the $i$th patron in the interaction. Furthermore, the beta-function can be calculated as \cite{Deppman:2019yno}, $\beta_g = -\big(g^3/16\pi^2\big) (q-1)^{-1}$,
scaling the properties of the QCD, one can obtain the beta-function of the QCD is  $\beta_{QCD}= -\big(g^3/ 16\pi^2 \big)  \Big[ \big(11 N_c/3\big)- \big(2N_f/3\big)\Big] $ \cite{Politzer:1974fr,Gross:1973id}. Thus, the non-extensive $q$ parameter remains a phenomenological parameter, with $q>1$, and is considered as a measure of non-additivity. It can be calculated in terms of $N_c$ and $N_f$ as \cite{Deppman:2019yno},
\bea
q= 1+ \frac{3}{11N_c -2N_f}.
\eea
The non-extensive $q$ parameter describes the gauge field interaction. Moreover, it can measure the sensitivity of the effective parton to its number degrees of freedom. In SU($3$) algebra, the non-extensive $q$ parameter is approximated to be $q\approx1.143$, which is in good agreement with experimental results giving $q=1.14\pm0.01$ \cite{Sena:2012ds,Marques:2015mwa}. Similarly, for SU($2$) algebra, this approximation gives $q\approx1.103$. In the present paper, we consider $q>1$, this because the non-extensive $q$ parameter has the typical value for high energy collisions in range $1 \leq q\leq1.2$ \cite{Marques:2015mwa, Cleymans:2013rfq, Li:2013kca, Azmi:2014dwa}.  In the large $N_c$-limit of QCD, where mesonic degrees of freedom become more prominent, the theory undergoes simplifications. In this context, the non-extensive $q$ parameter tends to approach one. This tendency arises from the substantial simplification of QCD dynamics in this limit, resulting in more additive behavior reminiscent of standard BG statistics, where $q$ equals one. To ensure that the $q$-exponential is always a positive real function, we assumed this condition$\big(1+(1-q) x \big)\geq0$, 
\bea
\exp_q{(x)}= \Big(1+(1-q)x\Big)^{\frac{1}{1-q}} = 0, \quad \quad \mathrm{for}\quad \Big(1+(1-q)x\Big) <0.
\eea
This condition is known as the Tsallis cut-ff prescription \cite{Rozynek:2015zca}. To avoid this condition, one can assume,
\bea
\exp_q{(x)}=  \left\{  
\begin{array}{cc} 
\Big(1+(1-q)x\Big)^{\frac{1}{1-q}}, \quad \quad \mathrm{for} \quad x\leq 0, \\ 
\Big(1+(q-1)x\Big)^{\frac{1}{q-1}}, \quad \quad \mathrm{for} \quad x> 0.
\end{array}
\right.
\eea 
Further details can be found in Ref. \cite{Rozynek:2015zca}. In this study, we investigate the chiral structure of the quark-hadron phase transition for different selections of the non-extensive $q$ parameter at various chemical potentials. 
\section{Results and Discussion \label{Results}}
This section explores the properties of quark-hadron phase structure using a PQM model with non-extensive Tsallis statistics. The results focus on various thermodynamic observables such as chiral condensates, deconfinement order parameters, subtracted condensates, and pseudo-critical temperatures, thermodynamic quantities, as well as fluctuations and correlations of conserved charges at finite temperature.
\subsection{Chiral Condensates and Deconfinement Order Parameters \label{OrderParameter}}
\begin{figure}[htb]
\centering{
\includegraphics[width= 17.5 cm, height=10.5 cm, angle=0  ]{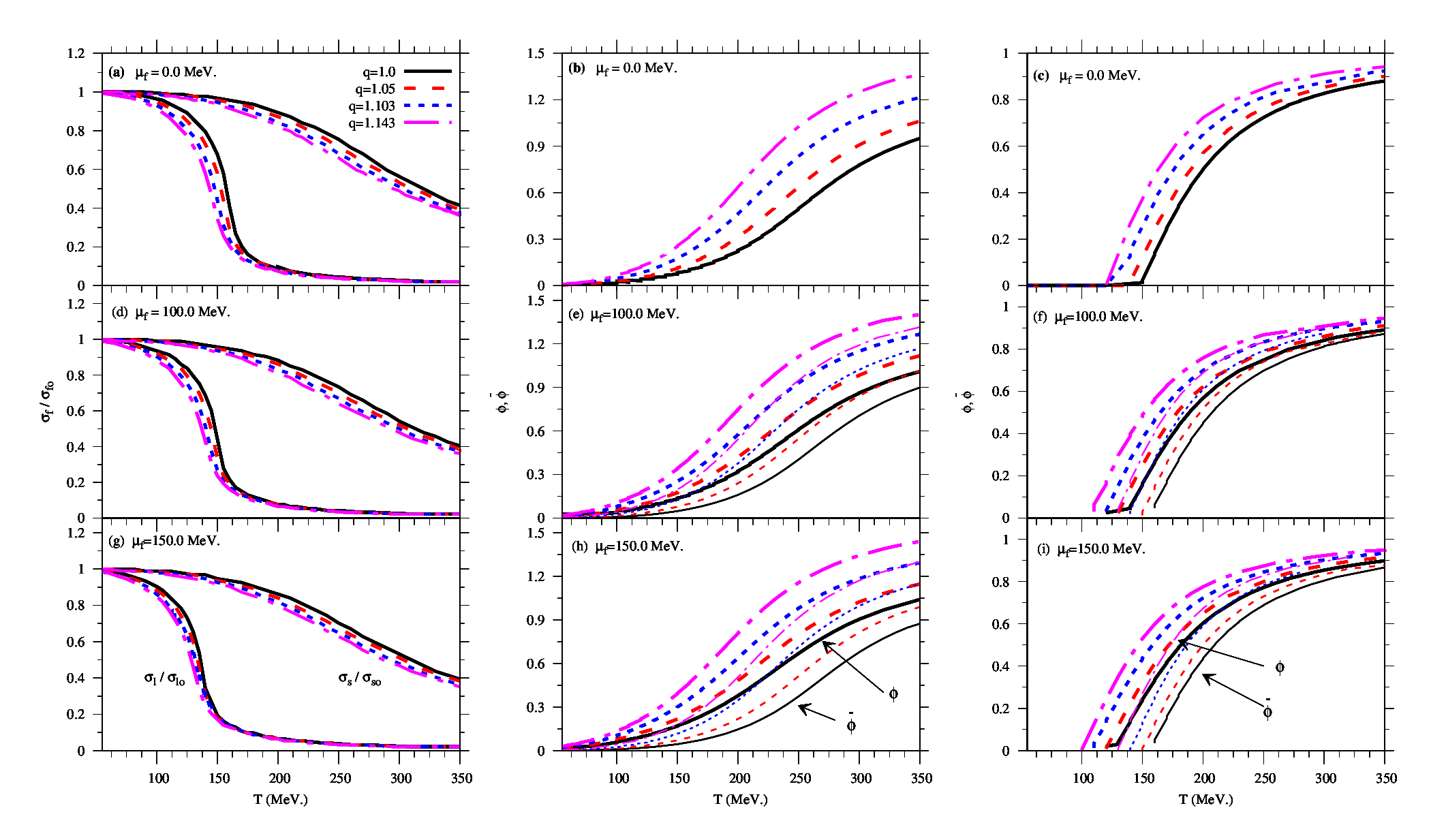}
\caption{Temperature dependence of the normalized (non)strange chiral condensates ($\sigma_\ell/\sigma_{\ell o}$) $\sigma_s/\sigma_{so}$ (left panel (a, d, g)) and the order parameter of Polyakov loop variables for the polynomial potential (middle panel (b, e, h))  and the Fukushima potential  (right panel (c, f, i)) at different selection of  chemical potentials $\mu_f = 0.0~$ (upper panel) $100.0$ (middle panel) and $150.0~$MeV (lower panel) are shown for different values of the non-extensive parameter $q=1.0$ (solid curves), $1.05$ (dashed curves), $1.103$ (dotted curves) and $1.143$ (dotted-dash curves), respectively.}
\label{Fig.orderParamterMU0}}  
\end{figure}

It is instructive to provide several calculations to demonstrate the influence of non-extensive effects on the chiral phase transition and the  deconfinement order parameters of the PQM model. In order to study the phase transition of a system, we can minimize the thermodynamic potential with respect to the corresponding mean field, as given by Eq. (\ref{saddleEq}). Fig. \ref{Fig.orderParamterMU0} illustrates the effects of the non-extensive $q$ parameter on: (left panels) the temperature dependence of the  quark constituents for normalized nonstrange and strange condensates ($\sigma_\ell/\sigma_{\ell o}$, $\sigma_s/\sigma_{so}$); (middle panels) the order parameter of the Polyakov-loop variables ($\phi$, $\bar{\phi}$) for the polynomial form of the Polyakov-loop potential; and the left panels are the same as middle panels, but for the Fukushima version, at different selection from the quark chemical potential $\mu_f = 0.0$ (upper panels), $100.0$ (middle panels) and $150.0~$MeV (right panels). For $q=1$  gives the usual BG statistics, it is quite similar to the case of $q\ne1.0$. At low temperature density the PQM order parameter keep the same values for different $q$. The left panel (a)  shows that the thermal behaviors of $\sigma_\ell/\sigma_{\ell o}$ and $\sigma_s/\sigma_{so}$ begins from the unity as their vacuum condensates $\sigma_{\ell o} =  92.5~$MeV and $\sigma_{so} =  94.2~$MeV. In this region, the impact of the non-extensive $q$ parameter appears to be negligible. However, as the non-extensive $q$ parameter increases, it appears to enhance the phase transition. The behavior of the chiral condensates is not a sharp phase transition but a smooth crossover. As $q$ increases, the transition occurs at a lower pseudo-critical temperature ($T_\chi$). The middle (b) and right (c) panels show the temperature dependence of the order parameters of Polyakov loop variables for the same values of non-extensive $q$ parameter for Polynomial and Fukushima potential, respectively. At $\mu_f=0.0~$MeV, we find that the mean values of $\phi$ and  $\bar{\phi}$ are identical, i.e. $\phi=\bar{\phi}$ for varying temperature and non-extensive $q$ parameter values. At low temperature regions, the confined hadron phase is dominant, and the effects of  non-extensive $q$ parameter are negligible. As the temperature increases, the confined hadron phase is transformed into the deconfined phase. The thermal evolution of the deconfinement phase transition appears to be smooth. We obtain the same conclusion regarding the non-extensive effects on chiral condensate, where the transition occurs at a smaller deconfined temperature $T_\phi$. The thermal behavior of the order parameters of Polyakov loop variables obtained in the polynomial potential is larger than that obtained in the Fukushima potential. At higher temperatures, the expectation values of the Polyakov loop variables for the polynomial potential (middle panels) can exceed the unity, which is considered unphysical and is related to the free energy of a static color charge. However, these results are in agreement with previous studies, as found in \cite{Fu:2007xc, Mao:2009aq, Schaefer:2009ui, Gupta:2009fg}. Furthermore, the Polyakov loop variables are sensitive to the geometry and topology of spacetime, which may exhibit behavior beyond the confined phase limits. In ansatz, the Fukushima potential replaces the higher-order terms of the $\phi$ and $\bar{\phi}$ polynomial potential with the logarithm of the Jacobi determinant, effectively avoiding divergences and expectation values larger than one. Furthermore, the PQM order parameters at $\mu_f =100.0$ (d), and $150.0~$MeV (g), it appears that the effect of chemical potential is negligible at low-temperature region. However, as the chemical potential increases, it enhances the phase transition. We observe that increasing the chemical potential helps reduce the impact of the non-extensive $q$ parameter on the phase transition, making it less severe than at low chemical potential. Also, we notice that the behavior of the $\sigma_s/\sigma_{so}$  remains longer than $\sigma_\ell/\sigma_{\ell o}$ which in turn is sensitive to $\mu_f$ and $q$ parameters. As the quark chemical potential $\mu_f$ is switched on, the thermal evolution of the order parameters for the deconfined phase transition becomes distinguishable, see  (middle panel (e, h)) and (right panel (f, i)). The mean field of Polyakov loop variables $\phi$ (thick curves) differs from $\bar{\phi}$ (thin curves), i.e., $\phi\ne\bar{\phi}$. It is obvious that the thermal behaviors of $\phi$ (thick curves) are evidently larger than their conjugate $\bar{\phi}$ as the quark chemical potential increases. The values of $\phi$ and their conjugate $\bar{\phi}$ increase as the non-extensive $q$ parameter is increased. The increase in $\mu_f$ and $q$ leads to a smooth increase in the Polyakov loop variables and enhances the deconfined phase transition. It is noteworthy that the impact of both $\mu_f$ and $q$ on the transition leads to the same conclusion: the deconfined transition occurs at a lower temperature $T_\phi$.

\subsection{Subtracted Condensates and Pseudo-Critical Temperatures \label{subtCond}}
As previously mentioned, the PQM model incorporates both strange and non-strange chiral condensates which are indicative of the chiral phase transitions. It would be valuable to demonstrate the non-extensive effects on the normalized net-difference or normalized subtracted condensate between the (non)strange chiral condensates as  \cite{Haas:2013qwp}
\bea
\Delta_{ls} (T,\;\mu_f)  =\frac{ \sigma_l (T,\;\mu_f) -(h_l/h_s) \;\sigma_s (T,\;\mu_f)}{\sigma_{\ell o} -(h_l/h_s) \;\sigma_{so} }, \label{subtracted}
\eea
where $h_l$ ($h_s$) are the explicit symmetry breaking parameters for non-strange (strange) quark sectors which can be obtained from the Dashen-Gell-Mann-Oakes-Renner (DGMOR) relations \cite{GellMann:1968rz, Dashen:1969eg}. 

\begin{figure}[htb]
\centering{
\includegraphics[width=15.5 cm, height=8.5 cm, angle=0  ]{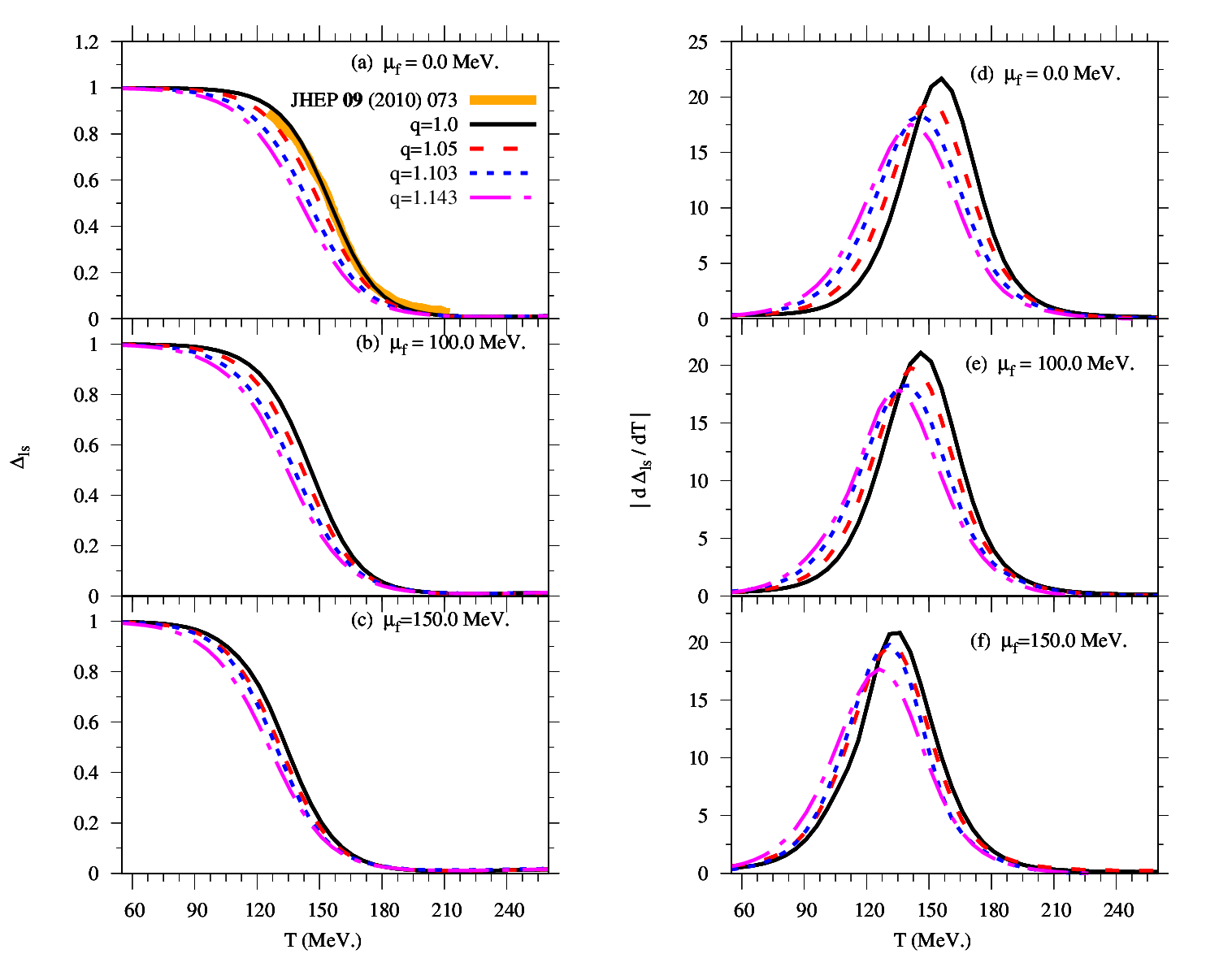}
\caption{The same as in Fig. \ref{Fig.orderParamterMU0}, but for the normalized net-difference the (non)strange chiral condensates $\Delta_{ls}$ (left panel) and the corresponding chiral susceptibility $\mid\partial \Delta_{ls}/\partial T\mid$ (right panel). The solid band gives the Continuum-limit of lattice QCD results \cite{Borsanyi:2010bp}. }
\label{Fig.subconddiffq}} 
\end{figure}

\begin{figure}[h]
\centering{
\includegraphics[width= 7.5 cm, height=9.5 cm, angle=-90  ]{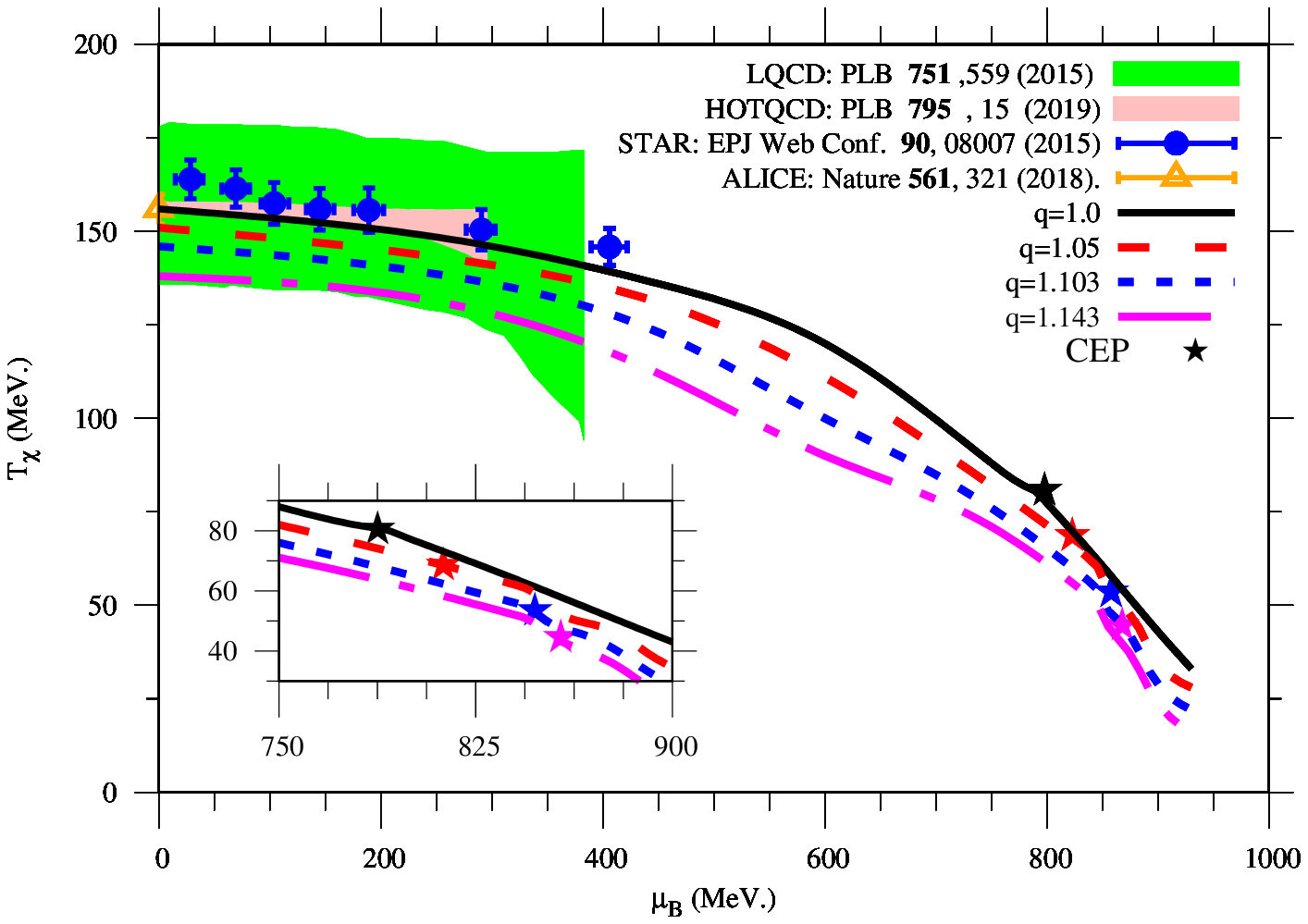}
\caption{The chiral phase diagram for different selections of non-extensive $q$ parameter. The PQM results are compared with lattice QCD simulations \cite{Bellwied:2015rza}  (solid wide band) and \cite{Bazavov:2018mes} (solid narrow band). Furthermore, these are confronted to the experimental results of ALICE  \cite{Andronic:2017pug} (open symbol) STAR experiments \cite{Das:2014qca} (closed symbols). The trajectory of CEP positions (inside-box) is shown as star symbols.}
\label{Fig.T-mu}} 
\end{figure}
Figure \ref{Fig.subconddiffq} illustrates the thermal behaviors of the normalized subtracted condensate between the (non)strange chiral condensates $\Delta_{ls}$ and the absolute value of the corresponding chiral susceptibility $\mid\partial \Delta_{ls}/\partial T\mid$ for different values of the non-extensive parameter $q=1.0$ (solid curves), $1.05$ (dashed curves), $1.103$ (dotted curves) and $1.143$ (dotted-dash curves) and different values of chemical potential $\mu_f= 0.0$ (upper panel), $100.0$ (middle panel) and $150.0~$MeV (lower panel). In the upper left-hand panel (a) of Fig. \ref{Fig.subconddiffq}, the PQM results for $q=1$ and $\mu_f=0.0~$MeV  are compared with the related lattice QCD simulation \cite{Borsanyi:2010bp}. There is good agreement, especially within the region of the phase transition. The thermal behavior of $\Delta_{ls}$ starts from unity and rapidly decreases as temperature increases until it becomes almost temperature independent at high temperatures. As $q-$parameter increases, the phase transition occurs at a lower pseudo-critical temperature ($T_\chi$). It is obvious that the non-extensive effects is negligible on the normalized subtracted condensate at low and high temperatures. 

The right-hand panel of Fig. \ref{Fig.subconddiffq} shows the first derivative of the subtracted condensate with respect to temperature in thermal and dense QCD medium for different values of the non-extensive $q$ parameter and varying chemical potentials $\mu_f$. The peak point indicates the pseudo-critical temperature at which the phase transition occurs. The peak position is shifted to lower temperatures as the non-extensive $q$ parameter increases. Furthermore, as the dense medium is released, the chemical potential repairs the phase transition towards lower temperatures. The obtained results can be used to map out the chiral phase diagram of PQM model. As discussed before, the pseudo-critical temperatures in PQM model, $T_{\chi}$, are estimated using different procedures. We can now map out $T_{\chi}$ at different baryon chemical potentials, $\mu_B=3\,\mu_f$, where the pseudo-critical temperature $T_\chi$ can be estimated as, 
\bea
T_\chi = \mbox{arg\; max}\; \Big[\frac{\partial \Delta_{ls}}{\partial T}\Big].
\eea 
  
Fig. \ref{Fig.T-mu} illustrates the non-extensive effects on the chiral phase diagram of the PQM model ($T_{\chi}-\mu_B$) plane. The PQM results at different values of $q=1.0$ (solid curves), $1.05$ (dashed curves), $1.103$ (dotted curves) and $1.143$ (dotted-dash curves) are compared with the  lattice QCD simulations \cite{Bellwied:2015rza} (solid wide band), \cite{Bazavov:2018mes} (solid narrow band) and also with the experimental observations in ALICE experiment at LHC (open symbols) \cite{Andronic:2017pug} and STAR experiment at RHIC (closed symbols) \cite{Das:2014qca}. The coordinates of ($T_{\chi},\,\mu_B$) were estimated by identifying the temperature at the peak point of the chiral susceptibility for various non-extensive $q$ parameter.  The PQM results for $q=1.0$ (solid curve) agree well with both lattice QCD calculations and experimental estimations. The calculations of the PQM model reveal that the chiral phase diagram, decrease with increasing the non-extensive $q$ parameter. As the non-extensive $q$ parameter increases, the pseudo-critical temperature decreases gradually by approximately $\sim 10\%$ compared to the previous value. The impact of non-extensive effects on the location of the CEPs (star symbols) is illustrated in Fig. \ref{Fig.T-mu} (inside-box). One can find how the positions of the CEPs vary with the non-extensive $q$  parameter. As $q$ increases, the CEPs shift rapidly towards higher chemical potential and lower temperature. This is because systems with fewer particles will experience a higher value of $q$, leading to phase transitions occurring at higher chemical potentials. The overlap of the curves forms an approximate straight line in the direction of lower temperature and higher chemical potential. In other words, the characteristic CEP should not only be viewed as a universal point but also as a critical region. Therefore, when analyzing experimental data, it is possible to encounter a critical region rather than a well-defined CEP. Moreover, the current search of the CEP position shows good agreement with other previous studies \cite{Rozynek:2009zh,Shen:2017etj,Zhao:2023xpj}.

\subsection{Bulk Thermodynamics \label{ThermoQ}}
Thermodynamic quantities plays role an important role in deriving  the features of hadron quark  matter in thermal and dense QCD medium. Furthermore, we can explore various thermodynamic observables from the thermodynamic pressure,  $p (T,\,\mu_f) = -\Omega (T,\,\mu_f)$ and the normalized interaction measure reads as,
\bea
\frac{I(T,\,\mu_f)}{T^4} =  \frac{\epsilon - 3p}{T^4} = T \; \frac{\partial}{\partial T} \Big(\frac{p}{T^4}\Big),
\eea
where $\epsilon (T,\,\mu_f)=  -p(T,\,\mu_f) + T\,s(T,\,\mu_f)$ is the energy density, while $s(T,\,\mu_f)=\partial p/\partial T$ is the entropy density. In Stefan-Boltzmann (SB) limit. The thermodynamic pressure  can be determined from the grand canonical partition function for $(N_c^2-1)$ massless gluons and $N_f$ massless quark flavors as an ideal gas \cite{Schaefer:2007pw},
\bea
\frac{p_{\tiny SB}}{T^4} = (N_c^2-1) \frac{\pi^2}{45} + N_c\,N_f \Big( \frac{7\pi^2}{180} + \frac{1}{6} \Big[ \frac{\mu_f}{T}\Big]^2 + \frac{1}{12 \pi^2} \Big[ \frac{\mu_f}{T}\Big]^4\Big). \label{PSBlimit}
\eea
The first term in rhs. of Eq. (\ref{PSBlimit}) refers to the gluonic contribution, the second term is the fermonic contribution and the other terms indicates contributions of ideal gas at finite chemical potential. Furthermore, we list the thermodynamic quantities in massless ideal gas at finite chemical potential \cite{Bhattacharyya:2016lrk} for  non-extensive Tsallis (TS) statistics as, 
\bea
\frac{p_{TS}}{T^4} &=& \frac{g}{6\pi^2}\; \frac{\big(1-\delta q\; \frac{\mu_f}{T} \big)^{3-\frac{1}{\delta q}} }{\big(1-\delta q\big) \big(\frac{1}{2}-\delta q\big) \big(\frac{1}{3}-\delta q\big)}, \nonumber \\
\frac{\epsilon_{TS}}{T^4} &=&  \frac{g}{2\pi^2}\;  \frac{\big(1-\delta q\; \frac{\mu_f}{T} \big)^{3-\frac{1}{\delta q}} }{\big(1-\delta q\big) \big(\frac{1}{2}-\delta q\big) \big(\frac{1}{3}-\delta q\big)},  \nonumber  \\
\frac{s_{TS}}{T^3} &=&  \frac{g}{6\pi^2}\;   \frac{ \big( 4-\; \frac{\mu_f}{T} -\; \delta q\;\frac{\mu_f}{T}  \big)  \big(1-\delta q\; \frac{\mu_f}{T} \big)^{2-\frac{1}{\delta q}} }{\big(1-\delta q\big) \big(\frac{1}{2}-\delta q\big) \big(\frac{1}{3}-\delta q\big)}. \label{TSEq}
\eea 
The degeneracy factor for massless ideal gas is defined as $g=g_{g} + (7/8) g_f$. For $N_f=3$ quark flavors, the QGP degree-of-freedom is $g=47.5$. Eqs. (\ref{TSEq}) are only valid for $0\leq (\delta q \equiv q-1 ) <1/3$ and $\delta q <T/\mu_f $ as required from the consistency of the framework.
\begin{figure}[htb]
\centering{
\includegraphics[width= 17.5 cm, height=5.5cm, angle=0]{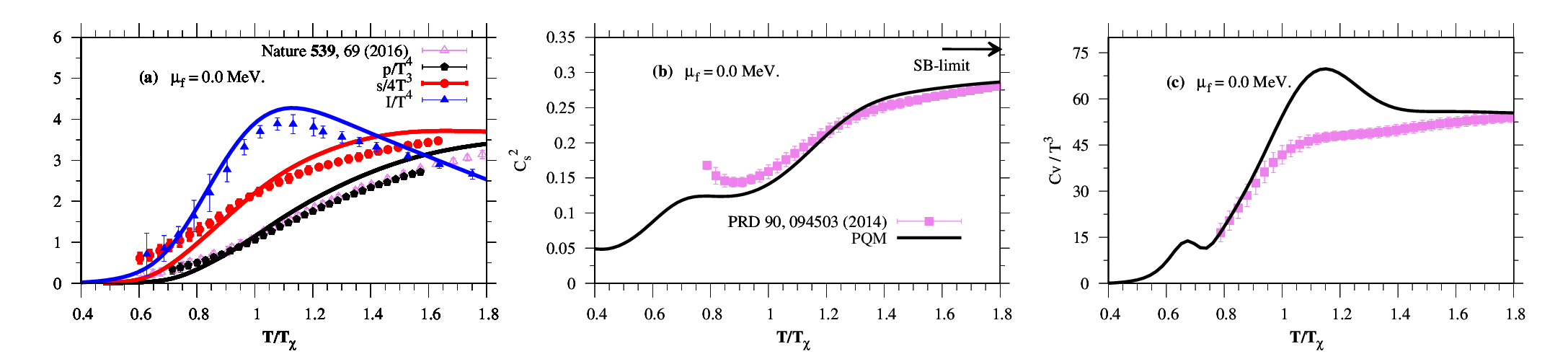}
\caption{(Color online) A comparison between the PQM thermodynamic quantities at $\mu_f=0.0~$MeV, normalized pressure, entropy and interaction measure (left-panel (a)), the speed of sound  squared (middle-panel (b)) and specific heat (right-panel (c)), with the lattice QCD simulations (symbols). The lattice QCD data are given by (open symbols)  \cite{Borsanyi:2016ksw}, (closed symbols) \cite{Borsanyi:2022qlh} and (closed squared-symbols) \cite{HotQCD:2014kol}.}
\label{Fig.thermoMU0}} 
\end{figure}

Figure \ref{Fig.thermoMU0} illustrates the temperature dependence of the various thermodynamic quantities such as normalized pressure $p/T^4$, entropy $s/T^3$ and  the interaction measure $I/T^4$ (left panel (a)), the speed of sound  squared $C_s^2$ (middle-panel (b)) and normalized specific heat $C_v/T^3$ (right panel (c)) obtained from the PQM model at $\mu_f=0.0$ MeV and $q=1.0$ in the usual BG statistics. The results are presented for Fukushima potential of Polyakov loop variables (solid curves). The PQM results exhibit good agreement with the lattice QCD calculations (symbols) \cite{Borsanyi:2016ksw,Borsanyi:2022qlh, HotQCD:2014kol}, particularly in the region of the phase transition. However, at low temperatures, the PQM results show a slight deviation from the lattice QCD calculations  \cite{Borsanyi:2016ksw,Borsanyi:2022qlh, HotQCD:2014kol}. On the other hand, at high temperatures, the PQM results align well with the lattice QCD calculations  \cite{Borsanyi:2016ksw,Borsanyi:2022qlh, HotQCD:2014kol}. The inclusion of the vector meson sector in the PQM model enhance the spatial resolution of these comparisons. It is worth emphasizing that the speed of sound  squared and the lattice QCD simulations appear to saturate below the SB-limit (middle panel (b)), near the pseudo-critical temperature,  it has a peak and then approaches the ideal gas value of SB-limit at high temperatures. It is worth noting that a peak appears in the vicinity of the pseudo-critical temperature for the normalized quantities $I/T^4$, $C_v/T^3$ and $C_s^2$ Subsequently, these values tend toward the massless ideal gas limit at higher temperatures.

\begin{figure}[htb]
\centering{
\includegraphics[width= 17.5 cm, height=5.5cm, angle=0  ]{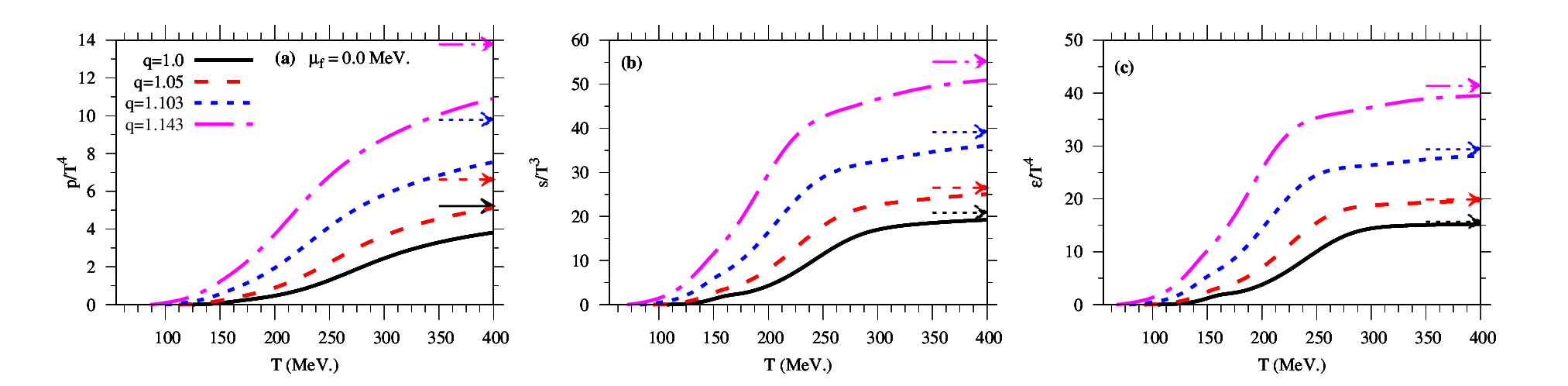}
\caption{Temperature dependence of various dimensionless thermodynamic quantities in the PQM model include the thermodynamic pressure $p/T^4$ (left panel (a)), entropy density $s/T^3$ (middle-panel (b)), and energy density $\epsilon/T^4$ (right panel (c)). Theses thermodynamic observables are given at  $\mu_f=0.0~$MeV and presented for various values of the non-extensive $q$ parameter: $q=1.0$ (solid curves), $q=1.05$ (dashed curves), $q=1.103$ (dotted curves), and $q=1.143$ (dotted-dash curves).}
\label{Fig.thermodiffq}} 
\end{figure}

\begin{figure}[htb]
\centering{
\includegraphics[width= 17.5 cm, height=5.5cm, angle=0  ]{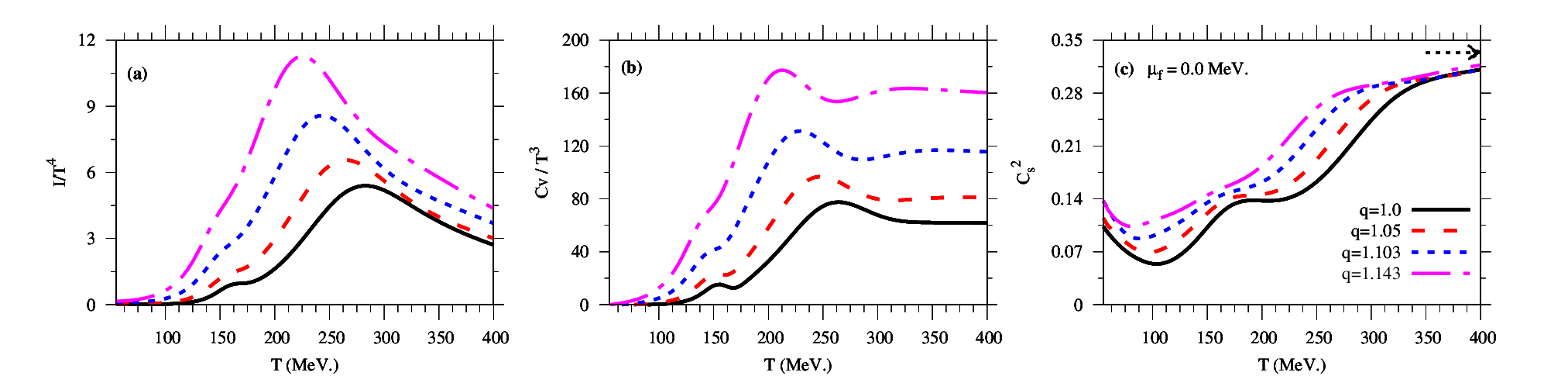}
\caption{The same as in Fig. \ref{Fig.thermodiffq}, but for other dimensionless thermodynamic observables including interaction measure $I/T^4$ (left panel (a)), specific heat $C_v/T^3$ (middle-panel (b)), and speed of sound  squared $C_s^2$ (right panel (c)).}
\label{Fig.thermo2diffq}} 
\end{figure}

Figures \ref{Fig.thermodiffq} and \ref{Fig.thermo2diffq} demonstrate the impact of the non-extensive $q$ parameter on the temperature dependence of various dimensionless thermodynamic quantities in the PQM model at $\mu_f=0.0~$MeV. In Fig. \ref{Fig.thermodiffq}, the thermal behaviors of the thermodynamic pressure $p/T^4$ (left panel (a)), entropy density $s/T^3$ (middle-panel (b)), and energy density $\epsilon/T^4$ (right panel (c)) are illustrated. Additionally, Fig. \ref{Fig.thermo2diffq} showcases other thermodynamic quantities: the interaction measure $I/T^4$ (left panel (a)), specific heat $C_v/T^3$ (middle-panel (b)), and speed of sound $C_s^2$ (right panel (c)), as functions of temperature $T$.

We observe a consistent increase in all thermodynamic quantities with an increasing non-extensive $q$ parameter. Notably, the influence of the non-extensive effect appears to be minimal in the low-temperature region, but becomes significant as it drives the phase transition from a smooth crossover to a prompt first-order phase transition. The thermal dependence of these quantities indicates their sensitivity to the deviation of the non-extensive effects, which reach a saturation value below the massless ideal gas limit (indicated by arrows). The thermodynamic observables are very sensitive to the deviation of the non-extensive $q$ parameter, they become larger in both phases of hadronic and QGP phases as $q$ increases. In light of this, the pseudo-critical temperature $T_{\chi}(\mu_B)$ seems not a universal constant, and decrease with increasing the non-extensive $q$ parameter. The incorporation of non-extensive statistics into the PQM thermodynamic  potential offers a replacement for the limit of a massless ideal gas, shifting towards the Tsallis limit instead of the SB limit. This suggests that the system, under Tsallis statistics at high temperatures, approaches the limit of a massless ideal gas at values exceeding the SB-limit. This limit is referred to as the $q$-dependent limit. It appears that the effects of non-extensive statistics on thermodynamic quantities persist even at high temperatures.

For relativistic heavy-ion collision, it is worth highlighting the correlation between the thermodynamic observables as the speed of sound squared as,
\bea
C_s^2 = \frac{\partial p}{\partial \epsilon}\mid_s = \frac{\partial p/\partial T \mid_V}{\partial \epsilon/\partial T \mid_V}= \frac{s}{C_v},
\eea
where $C_v = \partial \epsilon/\partial T \mid_V$ is the specific heat at constant volume. The speed of sound squared is also related to the interaction measure $I(T,\mu_f)$ as,
\bea
\frac{I(T,\mu_f)}{\epsilon(T,\mu_f)} = \frac{\epsilon(T,\mu_f) - 3 p(T,\mu_f)}{\epsilon(T,\mu_f)} \approx 1-3\; C_s^2, \label{DelTaCs2}
\eea
where the interaction measure can be related to the equation of state (EoS) $\epsilon(p)$. As stated in Eq. (\ref{DelTaCs2}), the minimum value of $C_s^2$ results in the maximum value of the ratio $I(T,\mu_f)/\epsilon(T,\mu_f)$ at low temperatures. Conversely, at high temperatures, as $C_s^2$ approaches the SB limit, the minimum value of the ratio $I(T,\mu_f)/\epsilon(T,\mu_f)$ is achieved. The right panel (c) of Fig. \ref{Fig.thermo2diffq} shows the  influence of the non-extensive $q$ parameter on the speed of sound squared as a function of $T$ at vanishing quark chemical potential. The thermal behavior of $C_s^2 $ is obtained from $(s/T^3)/(C_v/T^3)$. This panel demonstrates a relative sensitivity to the deviation of the non-extensive $q$ parameter in both the hadronic phase and the transition region, where it drives the phase transition as a smooth crossover. The system transitions more rapidly towards the QGP phase as the non-extensive $q$ parameter gradually increases. At high temperatures, the behavior of $C_s^2$ approaches the limit of a massless ideal gas, specifically $1/3$. The thermal behavior of $C_s^2$ exhibits an increasing trend, particularly in the hadronic phase, as the non-extensive $q$ parameter increases. However, this trend gradually disappears as temperatures rise. It is worth noting that the behaviors of $C_s^2$ for different $q$ parameters remain below the SB limit of $1/3$. This indicates that the behavior of $C_s^2$ remains below the SB limit and is consistent with the Tsallis limit at high temperatures. The speed of sound serves as a measure of the deviation from the equation of state of a massless ideal gas, which is represented by $\epsilon=3p$. 
\subsection{Fluctuations and Correlations of Conserved Charges \label{Fluctuations}}
Various fluctuations (diagonal) and correlations (off-diagonal) susceptibilities of different conserved quantum charges, $x=[B,\, Q,\,S]$, can be determined from the normalized thermodynamic pressure of PQM model as, 
\bea
\chi^{BQS}_{\alpha\,\beta\,\gamma} &=& \frac{\partial^{\alpha+\beta+\gamma}\; (p(T, \hat{\mu}_x)/T^4)}{(\partial \hat{\mu}_B)^\alpha\, (\partial \hat{\mu}_Q)^\beta\, (\partial \hat{\mu}_S)^\gamma\,}, \label{fluc_corr}
\eea
the superscripts $\alpha$, $\beta$, and $\gamma$ represent the orders of the derivatives. Eq. (\ref{fluc_corr}) illustrates that the fluctuations and correlations of conserved quantum charges can be obtained by taking derivatives of the dimensionless thermodynamic pressure with respect to the corresponding dimensionless chemical potential, as given in Eq. (\ref{muf-eqs}), where $\hat{\mu}_x = \mu_x/T$. Diagonal (off-diagonal) susceptibilities or moment products can be constructed, which can be linked to the measured multiplicities of produced particles, encompassing mean, variance, skewness, kurtosis, and other order cumulants \cite{Luo:2017faz}. This section focus on the kurtosis of net-quantum fluctuations. In particular, ratio of the second and fourth order cumulants as, $\kappa \sigma^2 =\chi^x_4/\chi^x_2$, which can be related to the measured multiplicities of produced particles \cite{Luo:2017faz}.

\begin{figure}[htb]
\centering{
\includegraphics[width= 17.5 cm, height=10.5 cm, angle=0  ]{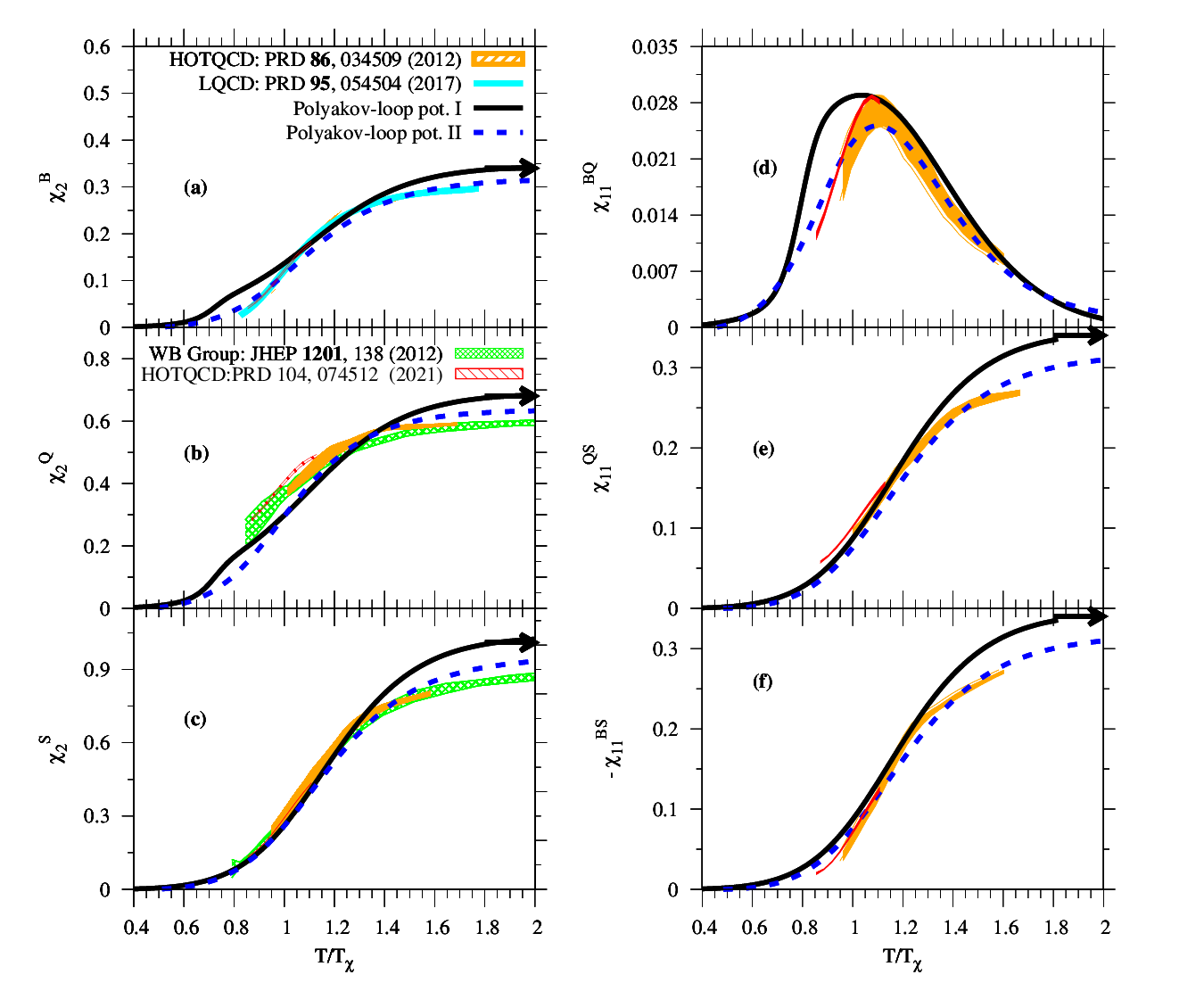}
\caption{A comparison of temperature dependence of the fluctuation (left panel) and correlations (right panel) of PQM model (solid and dashed curves) and lattice QCD simulations   (solid bands) at $\mu_f=0.0~$MeV.}
\label{Fig.Lattice2ndfluctuations}} 
\end{figure}

In this section, we investigate the thermal behaviors of diagonal (off-diagonal) susceptibilities of conserved charges up to the $8-$th order for non-extensive Tsallis statistics. Figure \ref{Fig.Lattice2ndfluctuations} presents a comparison of the second-order fluctuations (left panel) and correlations (right panel) with corresponding results obtained from lattice QCD calculations \cite{Borsanyi:2011sw, Bazavov:2012jq, Bazavov:2017dus} at $\mu_f=0.0~$MeV. The PQM model results are presented in two different forms: polynomial (solid curves) and Fukushima (dashed curves) potentials of the Polyakov loop variables, as functions of temperature $T$ at vanishing chemical potential. Remarkably, the PQM results demonstrate good agreement with the lattice QCD calculations \cite{Borsanyi:2011sw, Bazavov:2012jq, Bazavov:2017dus}, particularly in the region of the phase transition. Moreover, the PQM results employing the Fukushima potential (dashed curves) exhibit better alignment with the lattice QCD calculations \cite{Borsanyi:2011sw, Bazavov:2012jq, Bazavov:2017dus} compared to the polynomial potential (solid curves), we expect this result because the Fukushima potential incorporate higher-order Polyakov-loop variables and derived from strong coupling simulations. Furthermore, the inclusion of the vector meson sector in the PQM model enhances the spatial resolution of these comparisons, as discussed in our previous study on bulk thermodynamics.

In the right panel (d-f) of Fig. \ref{Fig.Lattice2ndfluctuations}, the thermal behavior of the correlations of conserved charges at $\mu_f=0.0~$MeV is depicted. The top panel (d) displays the correlation between the baryon number and the electric charge, denoted as $\chi^{BQ}_{11}$, as a function of $T$. At low temperatures, the correlation exhibits a gradual exponential increase as $T$ rises, eventually vanishing at high temperatures. This behavior is expected since the quarks become effectively massless in this temperature regime \cite{Borsanyi:2011bm, Cheng:2008zh}. The middle (e) and bottom (f) panels illustrate the correlations between strangeness and the electric charge, denoted as $\chi^{QS}_{11}$, and between strangeness and the baryon number, denoted as $\chi^{BS}_{11}$. These correlations are sensitive to variations in the strangeness degrees of freedom \cite{Koch:2005vg,Jeon:2000wg, Majumder:2006nq}. In the case of isospin symmetry considered in this study, these correlations are related to each other through the quadratic strangeness fluctuations $\chi^S_2=2\chi^{QS}_{11}-\chi^{BS}_{11}$ \cite{Jeon:2000wg, Majumder:2006nq}. At high temperatures, the correlations $\chi^{QS}_{11}$ and $\chi^{BS}_{11}$ become approximately equal, indicating a convergence of strangeness correlations.

The susceptibilities exhibit small values at low temperatures, indicating the dominance of the hadronic phase in this region. As the temperature enters the crossover region, the fluctuations rapidly increase. The temperature dependence of fluctuations and quark number multiplicities is significantly influenced by the chiral structure of the hadronic states. High temperatures give rise to a new state of matter characterized by massless quarks and gluons. As the system undergoes the transition to the deconfined phase, the fluctuations progressively increase and approaching a saturation value, slightly below the SB limit. 

\begin{figure}[htb]
\centering{
\includegraphics[width= 17.5 cm, height=10.5 cm, angle=0  ]{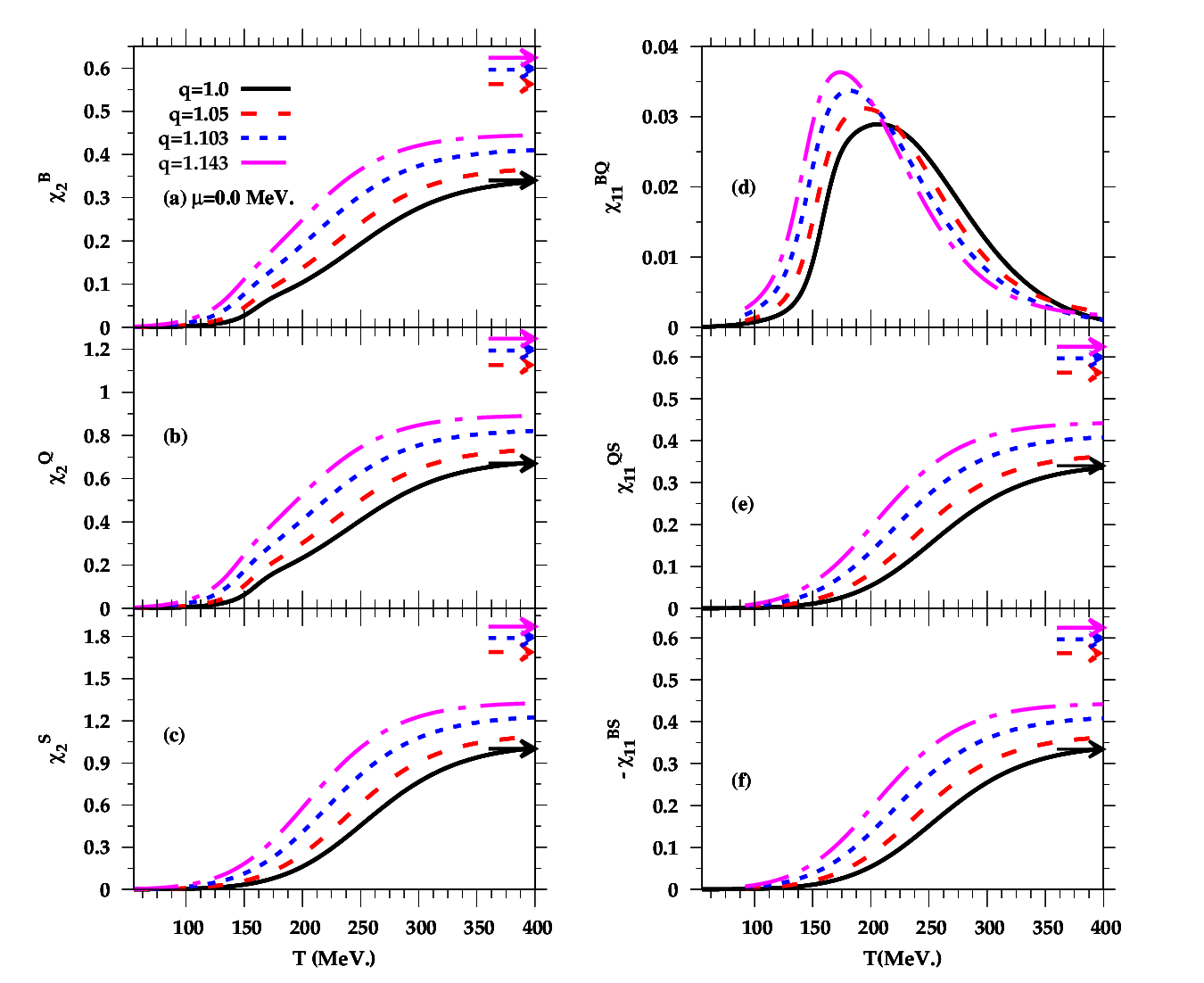}
\caption{The same as in Fig. \ref{Fig.Lattice2ndfluctuations}, but for different values of non-extensive $q$ parameter as $q=1.0$ (solid curves), $1.05$ (dashed curves), $1.103$ (dotted curves) and $1.143$ (dotted-dash curves).}
\label{Fig.2ndfluctuations}} 
\end{figure}

Figure \ref{Fig.2ndfluctuations} illustrates the influence of non-extensive Tsallis statistics on the second-order fluctuations (left panel (a-c)) and correlations (right panel (d-f)) in the PQM model as a function of temperature ($T$) at $\mu_f=0.0~$MeV. The results show that the second-order susceptibilities are sensitive to variations in the $q$ parameter, with all susceptibilities increasing as $q$ increases. At low temperatures, the non-extensive effects are negligible as the PQM order parameter remains unaffected by the non-extensive $q$ parameter. However, as the temperature increases, the impact of non-extensivity becomes more pronounced, enhancing the crossover until reaching a saturation value below the massless ideal gas limit (indicated by arrows) at high enough temperatures. Consequently, the pseudo-critical temperature decreases with increasing values of the non-extensive $q$ parameter.  By incorporating non-extensive Tsallis statistics into the PQM model, we introduce an alternative limit to the massless ideal gas, distinct from the SB limit. This alternative limit, known as the Tsallis limit, indicates that the system at high temperatures tends towards the massless ideal gas limit beyond the SB limit. This $q$-dependent limit has been discussed in our previous study on bulk thermodynamics.

In the left panel of Figure \ref{Fig.2ndfluctuations}, shows the correlation between the baryon number and the electric charge, denoted as $\chi^{BQ}{11}$ (top panel (d)), as a function of $T$ for different values of the non-extensive $q$ parameter. We observe a consistent increase in the thermal behavior as the non-extensive $q$ parameter increases. Additionally, the pseudo-critical temperature decreases with increasing values of the non-extensive $q$ parameter. At low temperatures, this thermal behavior exhibits an exponential increase as the temperature rises. Eventually, at high temperatures, the correlation approaches to vanish. Moreover, in the case of the correlations $\chi^{QS}_{11}$ (middle panel (e)) and $-\chi^{BS}_{11}$ (bottom panel (f)), they become approximately equal at high temperatures, indicating a convergence of strangeness correlations. These correlations reach a saturation value slightly below the limit of a massless ideal gas.

\begin{figure}[htb]
\centering{
\includegraphics[width= 18 cm, height=6cm, angle=0  ]{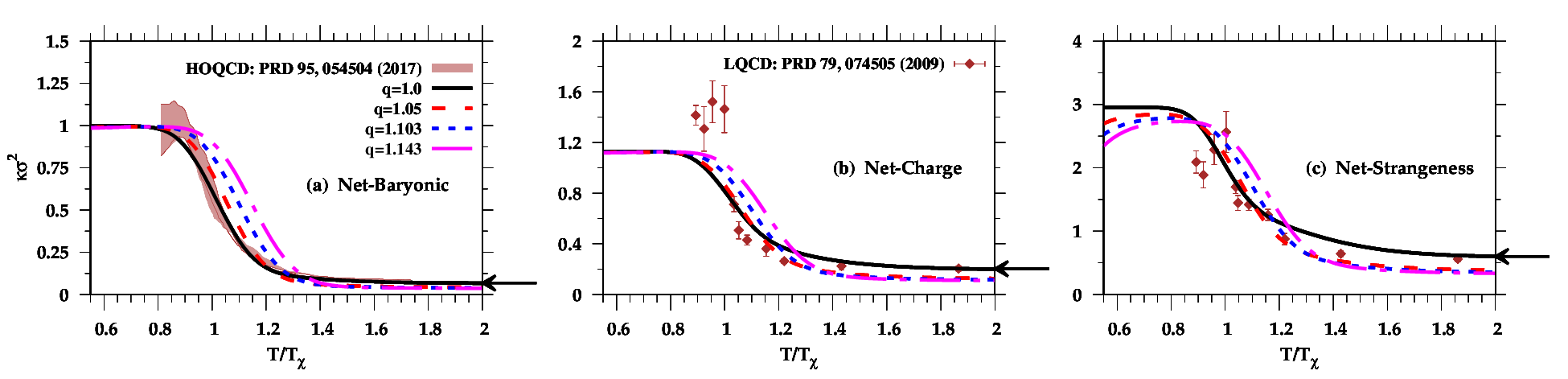}
\caption{The thermal dependence of the moment products $\kappa\sigma^2$ calculated from the PQM model for different values of non-extensive $q$ parameter as $q=1.0$ (solid curves), $1.05$ (dashed curves), $1.103$ (dotted curves) and $1.143$ (dotted-dash curves) for (a) net-baryon, (b) net-charge and (c) net-strangeness multiplicities compared with lattice QCD simulations (solid band) \cite{Bazavov:2017dus} and (closed symbols) \cite{Cheng:2008zh}. The solid arrows gives the associated SB limits. }
\label{Fig.fluctuationsX42q}} 
\end{figure}

\begin{figure}[htb]
\centering{
\includegraphics[width= 17 cm, height=10.5cm, angle=0  ]{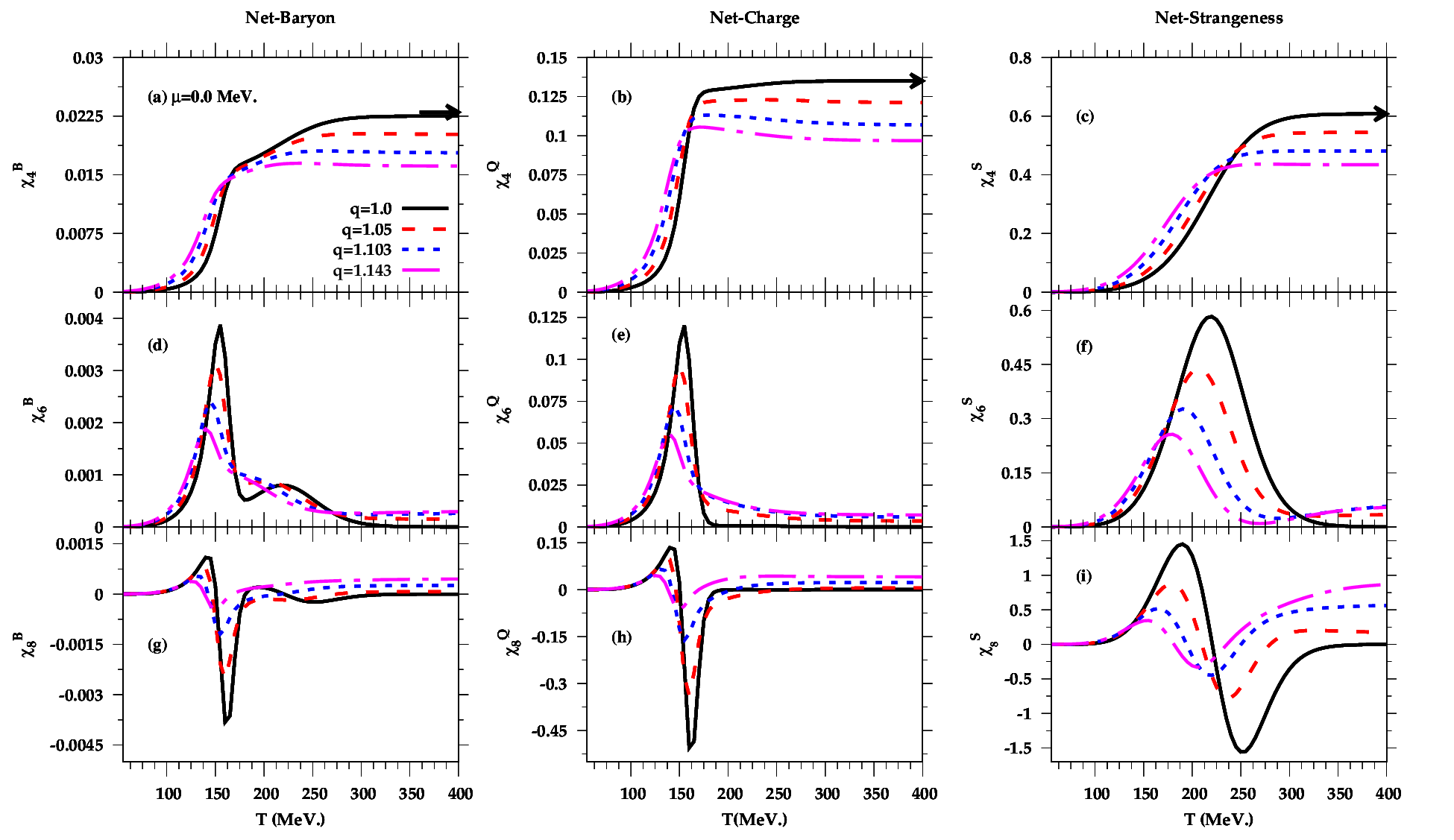}
\caption{The same as in Fig. \ref{Fig.2ndfluctuations}, but for the thermal behaviors of the  higher order fluctuations, $\chi^x_4, \;\chi^x_6$ and $\chi^x_8$, where $x$ refers to the net-baryon (left panel (a, d, c)), net-charge (middle panel (b, e, h)) and net-strangeness (right panel (c, f, i)) multiplicities.}
\label{Fig.Hmoments}} 
\end{figure}
 
The PQM Lagrangian  is constructed from the chiral Lagrangian coupled to the Polyakov loop variables. This implies that at low temperatures, the effective degrees of freedom of quark and anti-quark contributions are limited as long as the expectation values of the Polyakov loop variables are both zero. Consequently, excitations of one- and two-quark states are suppressed in the partition sum. Therefore, at low temperatures, the thermodynamic pressure of the quark and anti-quark contributions can be approximated, constituting what is known as the Boltzmann approximation \cite{Karsch:2003zq, Karsch:2003vd} as, 
\bea
\frac{p^{th}_{\psi\bar{\psi}}}{T^4} \sim \frac{2N_f}{27\pi^2} \; \Big(\frac{3m_f}{T}\Big)^2 \; K_2 \Big(\frac{3m_f}{T}\Big) \; Cosh \Big(\frac{3\mu_f}{T}\Big). \label{BoltzmannEq}
\eea
At high temperatures, it is expected that the susceptibilities will approach those of an ideal gas of quarks. Eq. \ref{BoltzmannEq} refers to the Boltzmann approximation is only valid if $\big(3m_f/T\big)>1$, which is satisfied for low temperatures $T<T_\chi$. A similar factorization occurs in HRG models as,
\bea
\frac{p^{th}_{\psi\bar{\psi}}}{T^4} \sim  F\Big(\frac{\mathcal{M}}{T}\Big) \; Cosh\Big(\frac{\mu_B}{T}\Big),
\eea
where $\mathcal{M}=3m_f$ and $\mu_B=3\mu_f$, the sum is taken over all known baryons and their resonances \cite{Ejiri:2005wq, Redlich:2004gp, Karsch:2003zq, Karsch:2003vd}. The kurtosis strongly depends on the effective degrees of freedom of quarks associated with the baryon number, (where $B=1$). For instance, at low temperatures, the kurtosis can be expressed as $\kappa\sigma^2 = (3B)^2 = 9$, a relationship also supported by observations in the HRG model\cite{Allton:2005gk,Ejiri:2005wq}. In contrast, in the pure chiral Quark-Meson model, where the relevant degrees of freedom consist of single quarks at all temperatures, the kurtosis at low temperatures converges to $\kappa\sigma^2 = 1$. It is evident that in an effective chiral model with quark degrees of freedom, the inclusion of the Polyakov loop is imperative for accurately describing the behavior of the kurtosis at low temperatures.

Figure \ref{Fig.fluctuationsX42q} displays the moment products $\kappa \sigma^2 = \chi^x_4/\chi^x_2$ computed using the PQM model at various values of the non-extensive parameter $q$ for (a) net-baryon, (b) net-charge, and (c) net-strangeness multiplicities. The results have sufficient agreement with lattice QCD simulations conducted by HOTQCD \cite{Bazavov:2017dus} (solid band) and other studies \cite{Cheng:2008zh} (closed symbols). The values of the kurtosis at high and low temperatures differ significantly, indicating variations near the phase transition. The substantial change in the kurtosis between the low and high-temperature limits primarily arises from the transformation of the relevant degrees of freedom carrying baryon number. These degrees of freedom transition from three-quark states to single quarks.

In the left-hand panel (a) of Fig. \ref{Fig.fluctuationsX42q}, the moment product $\kappa \sigma^2 = \chi^B_4/\chi^B_2$ starts at unity due to the inclusion of the Boltzmann approximation, Eq.(\ref{BoltzmannEq}), in the PQM Lagrangian within the confinement region. As the temperature $T$ increases, there is a rapid decrease in this behavior, attributed to the swift evolution of the quadratic susceptibility. At high temperatures, the kurtosis reaches a saturation level below the SB limit, $\chi^B_4/\chi^B_2|_{SB} = 2/3\pi^2$, and remains temperature-independent in the deconfined phase. The temperature dependent of kurtosis is very sensitive to variations in the $q$ parameter. At low temperatures, non-extensive effects are negligible. However, as the temperature rises, the influence of non-extensivity becomes increasingly pronounced, leading to an enhancement of the crossover until it reaches to saturated plateau below the SB limit.
  
The middle and right-hand panels of Fig. \ref{Fig.fluctuationsX42q} depict the moment product $\kappa \sigma^2$, following the same pattern as in the left-hand panel (a), but for (b) net-electric charge and (c) net-strangeness multiplicity, respectively. These PQM results are compared with lattice QCD calculations \cite{Cheng:2008zh} (closed symbols). Once again, at low temperatures, the influence of the Boltzmann approximation dominates during the confinement phase, and the effects of non-extensivity are negligible. As we approach the chiral transition, the kurtosis shows a strong dependence on the non-extensivity parameter, extending to higher temperatures where the susceptibilities approach those of an ideal gas, following the SB limit. There is a qualitative agreement, especially in phase transition region to high temperatures. The results approach the SB limits of $2/\pi^2$ and $6/\pi^2$ for net-electric charge and net-strangeness multiplicities, respectively.

Fig.  \ref{Fig.Hmoments} illustrates the impact of non-extensive $q$ effects on the higher order susceptibilities $\chi^x_n$ ($n=4, 6, 8$) for net-baryon $\chi^B_n$ (left panel (a, d, c)), net-charge $\chi^Q_n$ (middle panel (b, e, h)), and net-strangeness $\chi^S_n$ (right panel (c, f, i)) multiplicities as functions of temperature $T$. As the non-extensive $q$ parameter increases, the $n$-th order susceptibilities decrease. Notably, the most significant deviations from the non-extensive effects are observed in the higher order susceptibilities near the crossover region. Increasing the non-extensive $q$ parameter enhances all the susceptibilities, and the pseudo-critical temperature decreases accordingly. For $n=6$ and $8$, the higher order susceptibilities exhibit peaks around the temperature of the phase transition. As the non-extensive $q$ parameter increases, these susceptibilities display oscillations in higher harmonics.It is worth mentioning that net-strangeness exhibits higher fluctuations compared to net-baryon and net-charge.   Alternatively, it is crucial to consider the impact of non-extensive effects when comparing theoretically calculated higher order fluctuations with experimental measurements. Confidence in the measured system is necessary to account for these non-extensive influences. Moreover, the PQM results demonstrate notable characteristics, showing a smooth crossover between the hadronic and partonic phases.

\section{Conclusions \label{conclusion}} 
In conclusion, we have combined the PQM model and the non-extensive Tsallis statistics to study the hadron-quark transition to deviations of the BG statistics. The results highlight the influence of the non-extensive $q$ parameter on the phase transition and various thermodynamic quantities. At vanishing chemical potential, one can find that the pseudo-critical temperature of the chiral and deconfinement transition shifts towards a lower temperature as the non-extensive parameter $q$ increases. Moreover, at finite chemical potential and with an increasing $q$ parameter, the position of the pseudo-critical temperature moves towards lower temperatures and larger chemical potentials. This suggests that the pseudo-critical temperature is not a universal constant; its position in the ($T_{\chi}$-$\mu_B$) plane may depend on the non-extensive $q$ parameter. Furthermore, the search for the position CEP in relativistic heavy-ion collisions may be dependent on the parameters of non-extensive Tsallis statistics. It is worthwhile to mention that the CEP carries lower temperature and higher chemical potential with increasing $q$ parameter. One can consider that the location of the CEP is not a well-defined point but rather a critical region dependent on non-extensive effects. Notably, the impact of the non-extensive effect on PQM results seems small in the low-temperature region but becomes significant as it propels the phase transition from a smooth crossover to a prompt first-order phase transition. On other hand, with an increasing $q$ parameter, the phase transition occurs at a pseudo-critical temperature that shifts towards lower temperatures. This implies that variations in the $q$ parameter can show an additional phase transition. As the $q$ parameter increases, the system transitions to the deconfined phase more rapidly.

Furthermore, the agreement between the PQM model and lattice QCD calculations regarding various thermodynamic quantities, especially in the phase transition region, underscores the validity of the model. The non-extensive $q$ parameter affects on these quantities, leading to higher values as $q$ increases. The pronounced impact of non-extensive effects occurs around the phase transition, driving the system towards a prompt first-order phase transition from a smooth crossover. Additionally, the incorporation of non-extensive statistics deviates the system from the massless ideal gas limit, aligning it more closely with the Tsallis limit. On other words, the thermodynamic quantities at higher temperatures limit are no longer to SB limit but $q-$related Tsallis limit. The sensitivity of thermodynamic quantities to the $q$ parameter is evident, and the $C_s^2$ remains below the Stefan-Boltzmann limit, consistent with the Tsallis limit at high temperatures. At higher temperatures, the growth rate of the dimensionless entropy and specific heat tends to be the same. Therefore, the $C_s^2$ behavior at high temperature limit is not affected by the non-extensive $q$ parameter.  

Additionally, the investigation of diagonal and off-diagonal susceptibilities of conserved charges in the PQM model with non-extensive Tsallis statistics yields results that align well with lattice QCD calculations, particularly near the phase transition. The sensitivity of susceptibilities to variations in the non-extensive $q$ parameter is demonstrated, providing valuable insights into the behavior of the system close to the phase transition. These susceptibilities enhance the crossover trend as the $q$ parameter increases. Overall, this study significantly contributes to our understanding of non-extensive effects in the hadron-quark phase structure and opens up avenues for further research in this intriguing field. Furthermore, we studied the moments of the net-baryon, net-electric charge and net-strangeness multiplicity, $\kappa \sigma^2=\chi^X_4/\chi^X_2$, as a function $T$ at different $q$ parameter. It is worth noting that the system transitions more rapidly towards the deconfined phase, and it undergoes the massless ideal gas limit more quickly as the non-extensive $q$ parameter increases.

\section*{Acknowledgements}
The author expresses his gratitude to the anonymous referees for their invaluable feedback and constructive suggestions, which greatly enhanced the quality of this work.
%
\bibliographystyle{aip}
\bibliography{PQMREF} 

\end{document}